\documentclass{ifacconf}

\usepackage{graphicx}      
\usepackage{natbib}        
\usepackage{amsmath}        
\begin{document}
\begin{frontmatter}

\title{Fractional quantum numbers deduced from experimental ground state meson spectra\thanksref{footnoteinfo}} 

\thanks[footnoteinfo]{Work supported  by DFG under contract number xxx.}

\author[First]{Richard Herrmann} 

\address[First]{Berliner Ring 80, D-63303 Dreieich (e-mail: herrmann@gigahedron.com).}

\begin{abstract}                
Based on the Caputo definition of the fractional derivative the ground state spectra of mesons are classified as multiplets of the
fractional rotation group. The comparison with the experimental values leads to the conclusion, that quarks carry an up to now
unrevealed fractional multiplicative quantum number, which we call fractional hyper charge.
\end{abstract}

\begin{keyword}
Fractional calculus, fractional group theory, high energy physics, meson spectra.
\end{keyword}

\end{frontmatter}

\section{Introduction}
During the last 10 years the interest of physicists in non-local field theories has been steadily increasing. The main reason for this development is the expectation, that the use of these field theories will lead to a much more elegant and effective way of treating problems in particle and high-energy physics as it is possible up to now with local field theories. 

A particular subgroup  of  non-local field theories can be described with operators of fractional nature and is specified within the framework of fractional calculus [\cite{ol}]-[\cite{f44}]. 

Fractional calculus provides us with a set of axioms and methods to extend the concept of a derivative operator from integer order n to arbitrary order $\alpha$, where $\alpha$ is a real or complex value. 

A particular successful application within the framework of fractional calculus is fractional group theory, which extends  a classical symmetry group to the analogue fractional symmetry group. This  implies a generalization of the classical symmetry and allows for a broader description of natural phenomena.  

The investigation of the properties of the fractional rotation group alone has lead to a vast amount of intriguing and valuable results.

One reason for this  success  is the strategy, to interpret concrete experimental data and strictly verify the theoretical results with experimental findings.

Important achievements are:
\begin{itemize}
\item
Exact prediction of the masses of Y(4260) und X(4664) in  the charmonium-spectrum, which were verified experimentally afterwards [\cite{he05}].
\item
The first application of fractional calculus in nuclear theory: a successful description of ground state band excitations in even-even nuclei [\cite{he07}].
\item
A first derivation of the exact form of interacting fractional fields in lowest order of the coupling constant and a first 
application: a high precision mass formula for baryon masses [\cite{he08}].
\item
Deduction of the fractional dynamic symmetry group, which simultaneously describes the magic numbers in nuclei [\cite{he10a}] and metal clusters [\cite{he10b}].
\end{itemize}
In the following we will classify the complete set of available ground state meson spectra according to multiplets of the fractional rotation group ${_\textrm{\tiny{C}}}SO^\alpha(3)$ based on the Caputo definition of the fractional derivative. 
\section{Derivation of the mass formula}
In order to derive the mass formula we will use, we start with
the Caputo definition of a fractional derivative: [\cite{cap}]:
\begin{equation}
{_\textrm{\tiny{C}}}\partial_x^\alpha \, f(x) \!=\! 
\begin{cases}
({_\textrm{\tiny{C}}}\partial_{+}^\alpha f)(x) \!=\!  
\frac{1}{\Gamma(1 -\alpha)}   
     \int_{0}^x \!\!\!  d\xi \, (x-\xi)^{-\alpha} \frac{\partial}{\partial \xi}f(\xi)
&  \\
\qquad\qquad \textrm{$x \geq 0$} \\
   ({_\textrm{\tiny{C}}}\partial_{-}^\alpha f)(x) \!=\! 
\frac{1}{\Gamma(1 -\alpha)}  
     \int_x^0  \!\!\!  d\xi \, (\xi-x)^{-\alpha} \frac{\partial}{\partial \xi}f(\xi)
&\\
\qquad\qquad \textrm{$x<0$} \\
\end{cases}
\end{equation} 
Applied to  a function set $f(x)=x^{n \alpha}$ this leads to:
\begin{eqnarray}
{_\textrm{\tiny{C}}}\partial_x^{\alpha} \, x^{n \alpha} &=& 
\begin{cases}
\frac{\Gamma(1+n \alpha)}{\Gamma(1+(n-1)\alpha)} \, x^{(n-1)\alpha}&\textrm{$n > 0$} \nonumber \\
0&\textrm{$n=0$} 
\end{cases} \\
\label{cx}
&=& {_\textrm{\tiny{C}}}[n]  \, x^{(n-1)\alpha}
\end{eqnarray} 
where we have introduced the abbreviation $ {_\textrm{\tiny{C}}}[n]$. 

We now introduce the fractional angular momentum operators or generators of infinitesimal rotations 
in the $i,j$ plane on the $N$-dimensional Euclidean space:
 \begin{equation}
{_\textrm{\tiny{C}}}L_{ij}(\alpha)  =
i \hbar(x_i^\alpha {_\textrm{\tiny{C}}}\partial_j^\alpha-x_j^\alpha{_\textrm{\tiny{C}}}\partial_i^\alpha)
\end{equation}
which result from canonical quantization of the classical angular momentum definition.
The commutation relations of the fractional angular momentum operators are isomorphic to the fractional 
extension of the rotational group $SO(N)$:
\begin{eqnarray}
{_\textrm{\tiny{C}}} \, [L_{ij}(\alpha),L_{kl}(\alpha)] &=& i\hbar
{_\textrm{\tiny{C}}}{f_{ijkl}}^{mn}(\alpha){_\textrm{\tiny{C}}}L_{mn}(\alpha) \\
& &  \qquad\qquad\qquad i,j,k,l,m,n=1,2,..,N \nonumber
\end{eqnarray}
with structure coefficients ${_\textrm{\tiny{C}}}{f_{ijkl}}^{mn}(\alpha)$. Their explicit form depends on the 
function set the fractional
angular momentum operators act on and on the fractional derivative type used. 

\noindent
In the case $N=3$, according to the group chain
\begin{equation}
\label{gch}
{_\textrm{\tiny{C}}}SO^\alpha(3) \supset {_\textrm{\tiny{C}}}SO^\alpha(2)
\end{equation}
there are two Casimir-operators $\Lambda_i$, namely 
\begin{eqnarray}
\Lambda_2 &= &L_z(\alpha)  = {_\textrm{\tiny{C}}}L_{12}(\alpha)\\
\Lambda_3  &=&  L^2(\alpha) = {_\textrm{\tiny{C}}} L_{12}^2 (\alpha) +{_\textrm{\tiny{C}}} L_{13}^2 (\alpha) + {_\textrm{\tiny{C}}}L_{23}^2 (\alpha)
\end{eqnarray}
We 
introduce the two quantum numbers $L$ and $M$, which completely determine the eigen functions
$|LM\rangle$.
It follows
\begin{eqnarray}
\label{Lz}
{_\textrm{\tiny{C}}} \, L_z(\alpha) |LM \rangle  &=&
 \hbar \, \, {_\textrm{\tiny{C}}} \,[M]\, |LM\rangle \\ 
 & & \qquad\qquad\qquad M = 0,...,L \nonumber \\
\label{L2}
{_\textrm{\tiny{C}}} \, L^2(\alpha) |LM\rangle  &=&
 \hbar^2 {_\textrm{\tiny{C}}} \,[L]{_\textrm{\tiny{C}}} \,[L+1] \,|LM\rangle \\
 & & \qquad\qquad\qquad  L = 0,1,2,... \nonumber
\end{eqnarray}
Notice, that we have omitted the states with $M<0$. 

Now according to the group chain  (\ref{gch}) we associate a Hamiltonian $H$, which can now be written in terms of the Casimir operators of the algebras 
appearing in the
chain and can be analytically diagonalized in the corresponding basis. The Hamiltonian is explicitly given as:
\begin{equation}
\label{hamiltonSO3}
H = m_0 + \frac{a_0}{\hbar^2}  L^2+
    \frac{b_0}{\hbar}  L_z 
\end{equation}
with the free parameters $m_0,a_0,b_0$ measured in units [MeV] and the basis is $|LM\rangle$.

Therefore we obtain eigen values of the Hamiltonian as:  
\begin{equation}
\label{zlc}
E^\alpha_C = m_0 + a_0 {_\textrm{\tiny{C}}}[L]{_\textrm{\tiny{C}}}[L+1] + b_0{_\textrm{\tiny{C}}}[M]
\end{equation}
In the next section 
we will apply (\ref{zlc}) as a four parameter $\{ \alpha, m_0, a_0, b_0 \}$ chiral mass formula
to classify the multiplets of available meson spectra.

\section{Phenomenology of meson spectra}
With
(\ref{zlc}) 
we obtain a satisfying description of all mesonic ground state excitation spectra simultaneously  
if we associate the quantum numbers
$ \langle LM \rangle$ 
with the corresponding mesonic excitation states.

\begin{figure}[t]
\begin{center}
\includegraphics[width=84mm]{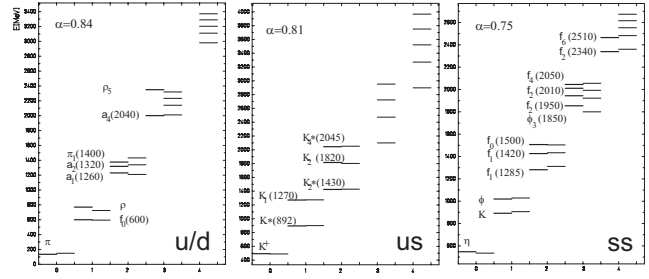}\\
\caption{
\label{fqs}
{
ground state meson excitation spectra for  $u \bar{u}$, $u \bar{s}$ (kaons) and $s \bar{s}$ (strangeonium). 
Thick lines represent experimental data [\cite{pdg}]. 
Names and experimental masses are given.
Thin lines denote the theoretical masses according to 
(\ref{zlc}).  
} }
\end{center}
\end{figure}

In a first simplified approach we will treat the up- and down-quarks as similar and will investigate the properties of mesons of the family
$\{u,s,c,b \}$. 
In figure
\ref{fqs}
the spectra for 
$u \bar{u}$, $u \bar{s}$ and
$s \bar{s}$ are presented. 
The vast amount of experimental data allows the determination of optimum parameter sets $\{ m_0, a_0, b_0  \}$ without difficulty.

The fitted optimum  parameter sets and the corresponding  root mean square error are given in table
 \ref{values}.

The predicted energies for the ground states  $|00\rangle$ cannot be described with standard phenomenological potential models [\cite{Ei75}] and are therefore an important indicator for the validity of the fractional concept.

Since we apply the Caputo fractional derivative, the ground state masses are directly determined by parameter $m_0$. It is remarkable, that we are able to  associate well known mesons successfully with the calculated ground state masses.

The ground state $|00\rangle$ of the 
$u \bar{u}$ system is predicted as $149 [\textrm{MeV}]$. 
Consequently we assign  a pion with experimentally deduced mass of 
$135 [\textrm{MeV}]$.
   
The ground state $|00\rangle$ of the
$u \bar{s}$ system is predicted as $489 [\textrm{MeV}]$, which very well corresponds to the experimental mass 
of the $K^+$ kaon with  $493 [\textrm{MeV}]$.
    
The ground state $|00\rangle$ of the
$s \bar{s}$ is predicted as  $534 [\textrm{MeV}]$. 
The lightest meson with a dominant  
$s \hat{s}$ content is the  $\eta$-meson 
with an experimental mass of
$547 [\textrm{MeV}]$.    

Therefore within the framework of our proposed simple model there is a remarkable coincidence of predicted ground state and observed experimental mass. This result strongly supports the validity of the proposed fractional concept.

\begin{figure}[t]
\begin{center}
\includegraphics[width=84mm]{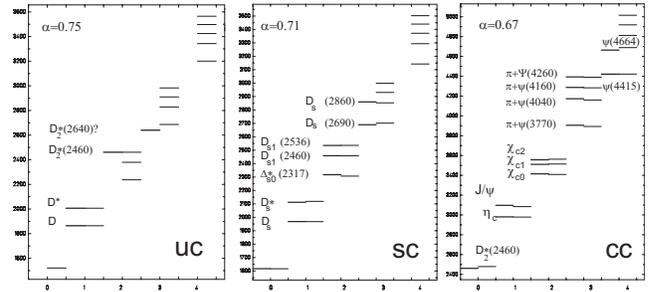}\\
\caption{
\label{fqc}
{
ground state meson excitation spectra for $u \bar{c}$ ($D$-mesons), $s \bar{c}$ ($D_s$-mesons)and $c \bar{c}$ (charmonium). 
Thick lines represent experimental data [\cite{pdg}]. 
Names and experimental masses are given.
Thin lines denote the theoretical masses according to 
(\ref{zlc}).  
} }
\end{center}
\end{figure}
\index{mesons!$D$}
\index{mesons!$D_s$}
\index{mesons!charmonium}
\begin{figure}[t]
\begin{center}
\includegraphics[width=84mm]{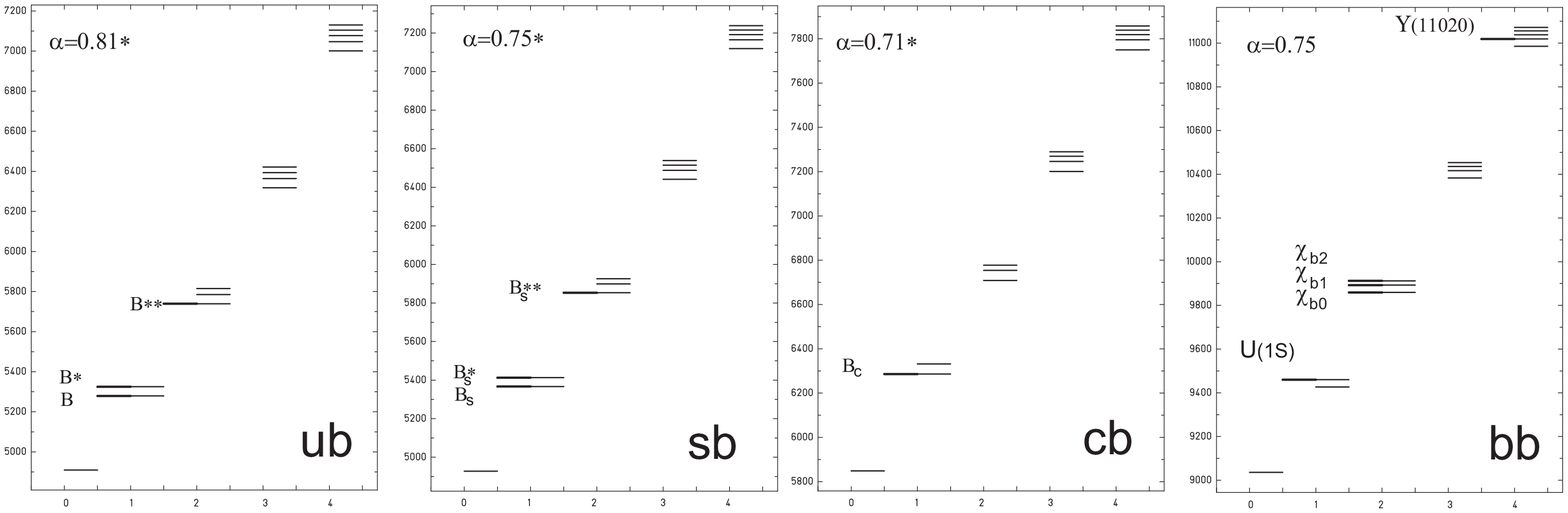}\\
\caption{
\label{fqb}
{
ground state meson excitation spectra for $u \bar{b}$ ($B$-mesons), $s \bar{b}$ ($B_s$-mesons), $c \bar{b}$ ($B_c$-mesons) and $b \bar{b}$ (bottomonium). 
Thick lines represent experimental data [\cite{pdg}]. 
Names and experimental masses are given.
Thin lines denote the theoretical masses according to 
(\ref{zlc}).  
} }
\end{center}
\end{figure}
In figure 
\ref{fqc} 
all spectra for mesons with $c$-quark content, in figure 
\ref{fqb} 
all spectra for mesons with $b$-quark content are collected. For these meson spectra for the predicted ground state $m_0$ we may associate an experimental value only in the case of the charmonium spectrum. For all other cases these particles are still missing in the experimental spectra.  

\begin{table*}[hb]
\begin{center}
\caption{
Optimum parameter sets for (\ref{zlc})
$\{\alpha, m_0, a_0, b_0\}$
and rms-error in percent for a fit with experimental meson spectra. For charmonium, we obtained 2 different $b_0$ values for energies below and above the $D-D*$ meson production threshold.
Since the experimental situation is not sufficient to determine all parameters simultaneously, in the lower part of this table all predicted parameters are labelled with 
(*).
 }
\label{values} 
\begin{tabular}{lllrrrr}
meson & name & $\alpha$ & $m_0$ & $a_0$ &$b_0$ & $\Delta$[$\%$] \\
\noalign{\smallskip}\hline\noalign{\smallskip}
$u\bar{u} \pm d\bar{d}$ & light unflavoured mesons & 0.842 & 149.26&   292.66& 137.85  &3.38\\
$u\bar{s} \pm s\bar{u}$ & kaons                             & 0.814 & 489.73&   281.56& 399.15  &0.48\\
$s\bar{s} $             & strangeonium                         & 0.751 & 534.96&   278.96& 132.57  &1.24\\
$c\bar{s} \pm s\bar{c}$ & $D_s$-mesons   & 0.716 &1617.06&   276.97& 156.85  &0.02\\
$c\bar{c} $             & charmonium    & 0.648 &2461.76&   452.59& 
$\begin{cases}
108.57 \!\!\!\!\!\!\!\!\!\!     \\
300.49 \!\!\!\!\!\!\!\!\!\!     \\
\end{cases}
$
 &0.18\\
$b\bar{b} $             & bottomonium   & 0.757 &9053.17&   269.67& 45.19  &0.06\\
\noalign{\smallskip}\hline\noalign{\smallskip}
$c\bar{u}, u\bar{c}$ & $D$-mesons       & 0.75* &1521.39&   258.10& 154.72  &0.0*\\
$u\bar{b}, b\bar{u}$ & $B$-mesons       & 0.80* &4907.71&   259.71&  49.39  &0.0*\\
$s\bar{b}, b\bar{s}$ & $B_s$-mesons     & 0.75* &4928.02&   326.98&  49.74  &0.0*\\
$c\bar{b}, b\bar{c}$ & $B_c$-mesons     & 0.71* &5848.63&   349.0*&  50.0*  &0.0*\\
\end{tabular}
\end{center}
\end{table*} 
In table \ref{values} we have listed the optimum parameter sets for all possible combinations of quark anti-quark pairs. 

Free parameters of the fractional model (\ref{zlc}) were determined by a fit with the experimental data. For light quarks there were no problems using a fit procedure since the experimental data base is broad enough. For heavy quarks we have only a limited number of experimental data. Consequently it is essential to deduce systematic trends in order to choose the parameter values despite the fact that there is only a small number of experimental data.

Using conventional phenomenological models a survey of strangeonium, charmonium and bottomonium states did not reveal a systematic trend for the parameter sets derived [\cite{hen79}]. The fundamental new aspect for an interpretation of meson spectra within the framework of a fractional rotation group is the fact, that systematic trends for the parameters may be deduced and may be used for a universal description of all possible mesonic ground state excitations.

For example from table \ref{values} we may deduce a systematic trend for the optimum fractional derivative parameter $\alpha$  for different mesonic systems. We may explain these values quantitatively, if we attribute to every quark and anti-quark respectively a multiplicative quantum number  
$\alpha_q$ (in analogy to  hyper-charge):
\begin{eqnarray}
\label{lalpha}
\alpha_u &=& \sqrt{{6 \over 7}} \\
\alpha_s &=& \sqrt{{6 \over 8}} \\
\alpha_c &=& \sqrt{{6 \over 9}} \\
\alpha_b &=& \sqrt{{6 \over 8}} 
\end{eqnarray}
with the requirement that $\alpha$ follows from the product
\begin{equation}
\label{xalpha}
\alpha = \alpha_{q_1 \bar{q}_2} = \alpha_{q_1} \alpha_{\bar{q}_2}
\end{equation}
As a consequence, using (\ref{xalpha}) we may make predictions for quark-anti-quark combinations, where the experimental basis is yet too small to determine all parameters of the model (\ref{zlc}) simultaneously.
These predicted  $\alpha$-values are labelled with (*) in table  \ref{values}.

A comparison of the optimum parameter sets for different quark anti-quark combinations reveals some common aspects:

For all mesons except charmonium the parameter $a_0$ is of similar magnitude of order
$280 [\textrm{MeV}]$ with a weak tendency to higher values for heavy mesons.

In addition, the value for the fractional magnetic field strength $b_0$ for all mesons except kaons is about
$150 \,[\textrm{MeV}]$. Furthermore there seems to exist a saturation limit for the field strength
$b_0= B/m_q$ which is reached for charmonium. 
For bottomonium which is about three times heavier, the calculated $b_0$ 
turns out to be $50 \,[\textrm{MeV}]$, which is $1/3$ of the charmonium value. Most probably an additional enhancement of the field strength could exceed the critical energy density for the creation of a light quark anti-quark pair e.g. a pion.    

Consequently in view of a fractional mass formula  (\ref{zlc}) the different meson spectra have a great deal in common which allows for an estimate of the spectrum of 
$c \bar{b}$-mesons, where up to now, only the $B_c$-meson has been verified experimentally. In table \ref{values} we have indicated the values interpolated with (*).

To summarize our results:
Based on the fractional mass formula  (\ref{zlc}) 
we found a general systematic in the ground state excitation spectra of mesons. Consequently we were able to classify all ground state excitation states for all possible 
mesons up to the bottom-quark as multiplets of the fractional rotation group $\textrm{SO}^\alpha(3)$.
   
\section{Internal structure of quarks}
Finally we may speculate about an internal structure of quarks.
We have calculated the ground state masses $m_0$ for all possible excitation spectra of mesons.

Let us assume that $m_0$ is the sum of the masses of two hypothetical constituent quarks
($Q_i$): 
\index{constituent quarks}
\begin{equation}
\label{m0}
m_0(Q_1,Q_2) = m_{Q_1} + m_{Q_2}
\end{equation}
We perform a fit procedure of parameter $m_0$
from table  \ref{values} according to  (\ref{m0}) and obtain:
\begin{eqnarray} 
Q_u &\approx&   75 [\textrm{MeV}] \\
Q_s &\approx&  302 [\textrm{MeV}] \\
Q_c &\approx& 1210 [\textrm{MeV}] \\
Q_b &\approx& 4800 [\textrm{MeV}] 
\end{eqnarray} 
A remarkable coincidence is the fact that
 $Q_s \approx 4 Q_u$, $Q_c \approx 4 Q_s$, $Q_b \approx 4 Q_c$, the mass of every constituent quark is four times its predecessor.
We obtain a rule
\begin{equation}
Q_j = m_u\, 4^j, \qquad j=1,2...
\end{equation}
with $m_u=18.92 [\textrm{MeV}]$. 

From this result we may speculate, that quarks exhibit an internal structure.
Four constituents with mass $m_u$
constitute an up-constituent quark, 16 built up a strange-constituent quark etc. Therefore we naively guess, that this is a hint for an internal    
$\textrm{SU}(4)$-symmetry. 

Furthermore we can make predictions for heavier quarkonia. We obtain
\begin{eqnarray}
b^* \bar{b}^*(j=5) &\approx& 38.7 \pm 5.0[\textrm{GeV}], \quad \alpha = \sqrt{{6 \over 7}}\\
b^* \bar{b}^*(j=6) &\approx& 155 \pm 20[\textrm{GeV}],  \quad \alpha = 1
\end{eqnarray}
So we are left with the problem, why these quarkonia have not been observed yet. 

It should be mentioned, that confinement is a natural property of free solutions  of a fractional wave equation with $\alpha < 1$. Therefore the constituents of the $j=6$ quarkonia are predicted to be quarks, which are not confined. As a consequence they are  free particles. It is at least remarkable, that indeed there are direct observations of top-events near $171.3 [\textrm{GeV}]$. Therefore within the framework of fractional calculus, the top-quark could be interpreted as a $j=6$ constituent quarkonium state.  

We have shown, that the full variety of mesonic excitations may be interpreted as a fractional rotation spectrum, if we apply the Caputo definition of the fractional derivative. The fractional parameter $\alpha$, deduced from experiment, may be interpreted as a multiplicative fractional quantum number, which may  be assigned to a specific quark flavour in analogy to the well known hyper charge.

Since in [\cite{he08}] we demonstrated, that the ground state baryon spectra may be interpreted as multiplets of the fractional rotation group based in the Riemann derivative definition,
 the only difference between baryons and mesons within the framework of fractional calculus is the presence and absence of a zero point energy contribution to the fractional rotational energy respectively. 

This is a first indication, that Caputo- and Riemann fractional derivative describe distinct fundamental physical properties, which are directly related to  fermionic and bosonic systems respectively.

\begin{ack}
We thank A. Friedrich and G. Plunien from TU Dresden, Germany for useful discussions. This work was supported by DFG under contract number xxxxxx.
\end{ack}

\end{document}